\documentclass{article}
\pdfoutput=1
\usepackage{arxiv}

\usepackage[utf8]{inputenc} % allow utf-8 input
\usepackage[T1]{fontenc}    % use 8-bit T1 fonts
\usepackage{hyperref}       % hyperlinks
\usepackage{url}            % simple URL typesetting
\usepackage{booktabs}       % professional-quality tables
\usepackage{amsfonts}       % blackboard math symbols
\usepackage{nicefrac}       % compact symbols for 1/2, etc.
\usepackage{microtype}      % microtypography
\usepackage{lipsum}		% Can be removed after putting your text content
\usepackage{graphicx}
\usepackage{natbib}
\usepackage{doi}
\usepackage{cite}
\usepackage{amsmath}
\usepackage{MnSymbol}
\usepackage{textgreek}

\graphicspath{{./figures/}}

\title{Palimpsest Memories Stored in Memristive Synapses}

\author{ \href{https://orcid.org/0000-0002-2002-2237}{\includegraphics[scale=0.06]{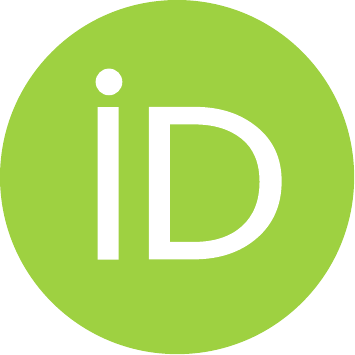}\hspace{1mm}Christos Giotis}\thanks{This work has been supported by the EPSRC EP/R024642/1 programme grant.}\thanks{Corresponding author} \\
	Centre for Electronics Frontiers\\
	Department of Electronics and Computer Science\\
	University of Southampton\\
	Southampton, UK, SO17 1BJ \\
	\texttt{c.giotis@soton.ac.uk} \\
	\And
	\href{https://orcid.org/0000-0002-8034-2398}{\includegraphics[scale=0.06]{orcid.pdf}\hspace{1mm}Alexander Serb} \\
	Centre for Electronics Frontiers\\
	Department of Electronics and Computer Science\\
	University of Southampton\\
	Southampton, UK, SO17 1BJ \\
	\texttt{a.serb@soton.ac.uk} \\
	\And
	\href{https://orcid.org/0000-0002-4931-8816}{\includegraphics[scale=0.06]{orcid.pdf}\hspace{1mm}Vasileios Manouras} \\
	Centre for Electronics Frontiers\\
	Department of Electronics and Computer Science\\
	University of Southampton\\
	Southampton, UK, SO17 1BJ \\
	\texttt{v.manouras@soton.ac.uk} \\
	\And
	\href{https://orcid.org/0000-0002-0833-6209}{\includegraphics[scale=0.06]{orcid.pdf}\hspace{1mm}Spyros Stathopoulos} \\
	Centre for Electronics Frontiers\\
	Department of Electronics and Computer Science\\
	University of Southampton\\
	Southampton, UK, SO17 1BJ \\
	\texttt{s.stathopoulos@soton.ac.uk} \\
	\And
	\href{https://orcid.org/0000-0002-6267-6909}{\includegraphics[scale=0.06]{orcid.pdf}\hspace{1mm}Themis Prodromakis} \\
	Centre for Electronics Frontiers\\
	Department of Electronics and Computer Science\\
	University of Southampton\\
	Southampton, UK, SO17 1BJ \\
	\texttt{t.prodromakis@soton.ac.uk} \\
}

\hypersetup{
pdftitle={Palimpsest Memories Stored in Memristive Synapses},
pdfauthor={Christos ~Giotis, Alexander ~Serb, Vasileios ~Manouras, Spyros ~Stathopoulos, Themis ~Prodromakis},
pdfkeywords={Palimpsest, Artificial memories, Memory consolidation, Memristors, Continuously-on learning, Artificial intelligence},
}

\begin{document}
\maketitle

\begin{abstract}
Biological synapses store multiple memories on top of each other in a palimpsest fashion and at different timescales. Palimpsest consolidation is facilitated by the interaction of hidden biochemical processes that govern synaptic efficacy during varying lifetimes. This arrangement allows idle memories to be temporarily overwritten without being forgotten, in favour of new memories utilised in the short-term. While embedded artificial intelligence can greatly benefit from such functionality, a practical demonstration in hardware is still missing. Here, we show how the intrinsic properties of metal-oxide volatile memristors emulate the hidden processes that support biological palimpsest consolidation. Our memristive synapses exhibit an expanded doubled capacity which can protect a consolidated long-term memory while up to hundreds of uncorrelated short-term memories temporarily overwrite it. The synapses can also implement familiarity detection of previously forgotten memories. Crucially, palimpsest operation is achieved automatically and without the need for specialised instructions. We further demonstrate a practical adaptation of this technology in the context of image vision. This showcases the use of emerging memory technologies to efficiently expand the capacity of artificial intelligence hardware towards more generalised learning memories.

\end{abstract}

% keywords can be removed

\section{Introduction}

While neural networks in the cerebral cortex employ an estimated $10^{13} - 10^{14}$ synapses to facilitate a plethora of cognitive abilities \citep{Shepherd2004, Koch1998}, their engineered counterparts require equivalent numbers of trainable parameters for a far narrower application spectrum \citep{LeCun2015, Xu2018}. One candidate for explaining this discrepancy in learning capacity between biological and artificial intelligence (AI), suggests that synapses are able to consolidate multiple memories which can be revealed at different timescales – much like a palimpsest \citep{Benna2016}. Synapses can remember long-term plasticity events, namely potentiation (LTP) and depression (LTD), while expressing altered states in the short-term \citep{Abraham2008}. This temporal partition enables the brain to use the same resources for multiple computation processes. The adoption of such flexibility by neuromorphic hardware is therefore a critical milestone towards the integration of AI in a wider range of on-the-edge, continuously-on learning systems.

Palimpsest storage is realised biologically via the bidirectional interaction of hidden biochemical processes affecting the manifestation of synaptic efficacy at different timescales, after each memory modification. These processes are characterised by their own degrees of plasticity (i.e. learning rates) and lifetimes (i.e. ‘forgetting time constants’). Sparsely presented memories induce fast changes in synaptic efficacy, but these quickly decay to reveal older but more persistent memories, that have successfully affected less plastic, but more long-lasting processes. The coexistence of such processes allows synapses to be both plastic in the short-term, enabling new memories to be written easily, and rigid in the long-term, thus preserving old memories of validated significance. 

The flexibility promised by dynamic memory consolidation has naturally attracted the attention of AI hardware design and in particular that of memristive technologies which have already showcased their potential in numerous neuromorphic applications \citep{Stathopoulos2017, Serb2016, Gupta2016, Berdan2016, Yoon2018, Boybat2021}. Memristors have been used to implement metaplasticity, i.e tuning of the learning rate \citep{Abraham1996, Fusi2005}, on CMOS-based artificial synapses in spiking neural networks (SNNs) \citep{Brivio2019, Demirag2021}. In a similar vein, non-volatile resistive RAM (RRAM) synapses use explicitly modulated bias voltage to tune their switching (i.e. learning) rate \citep{Zhu2017,Wu2018,Cheng2018,Liu2018,Lee2018}. Volatile RRAM has also demonstrated short to long-term memory transition where repeated presentations of the same memory induce longer changes in synaptic states, albeit being irreversible and unidirectional \citep{Chang2011, Tan2016,Ohno2011}. Finally, simulated artificial neural networks (ANNs) based on metaplasticity principles have demonstrated an increase in specialised learning capacity \citep{Laborieux2021}. While the authors comment that palimpsest capabilities can expand capacity towards uncorrelated memories, currently neither commercially available nor emerging memristive technologies have exhibited the necessary properties required for dynamic memory consolidation.

In this work the characteristics of RRAM volatility \citep{Giotis2020, Giotis2020b} are exploited to emulate the function of the hidden biochemical processes that enable palimpsest consolidation. We harness the bidirectional volatile and non-volatile responses of RRAM devices to practically realise two consolidation timescales in one device, effectively storing competing binary states in a single synapse with doubled short- and long-term memory capacity. Then, we upscale this principle to consolidate memory traces at variable consolidation intensity. Our technology can protect a strong memory in its long-term storage while allowing multiple short-term signals to take over the short-term memory fleetingly and then quickly decay. Simple metaplasticity is also realised as a natural subset behaviour when a given memory is consolidated consistently. Lastly, we show how this palimpsest memory can operate in a visual system where it also boasts unsupervised denoising abilities. Our memory system is unique in a number of key attributes. First, its expanded capacity is independent of the correlation between memories presented to it and even performs under fully destructive interference. Moreover, it can automatically sacrifice palimpsest capabilities for even stronger memory protection. These features unfold naturally as a result of single memristive device properties and do not require special biasing regimes or otherwise increased operational complexity.

\section{Results}
\subsection{Volatile memristors as candidates for palimpsest consolidation}

\begin{figure}[ht]
    \centering
    \includegraphics{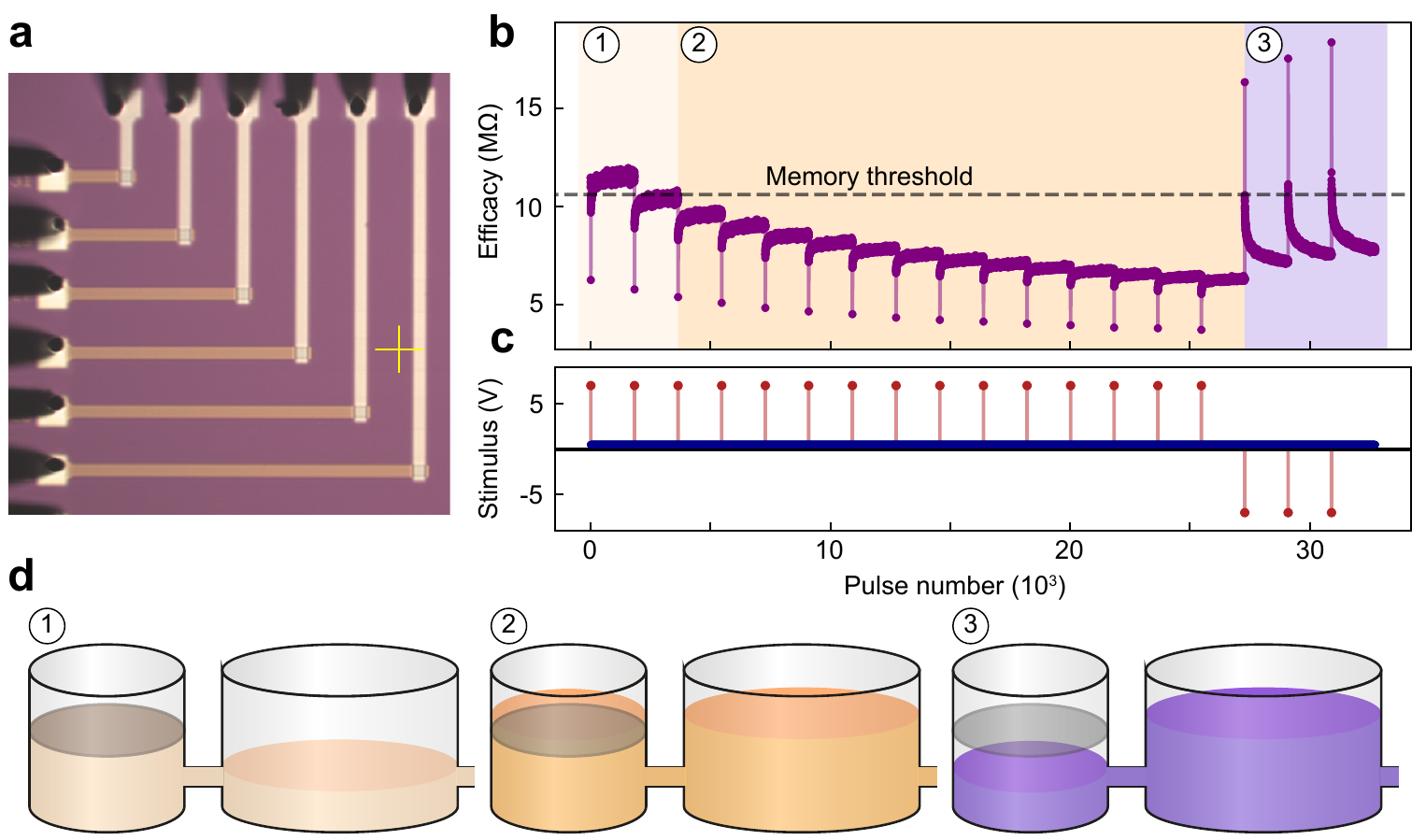}
    \caption{\textbf{Demonstration of a memristive synapse.} (\textbf{a}) Schematic operation of a volatile memristor. Plasticity events cause a pronounced and rapid change in synaptic efficacy observed on a fast timescale (short-term memory) and a smaller but more stable change on a slow timescale (long-term memory) (\textbf{b-c}) The change in synaptic efficacy over time is shown in \textbf{b} following the memory pattern applied in \textbf{c}. Two binary states are consolidated at separate timescales over three consolidation stages.  Stage 1:  The rigid component of the synapse lies close to binary threshold.  Potentiation events cause plastic storage of S\textsubscript{1}, which becomes noisy over time.  Stage 2:  Progressive potentiation events induce more rigid consolidation of S\textsubscript{1} as synaptic efficacy remains below the memory threshold.  Stage 3:  Depression events induce a plastic STM storage of S\textsubscript{0}.  Then, the synapse reinstates S\textsubscript{1} which has been consolidated on the slow/rigid timescale. (\textbf{d}) Schematic equivalence of the three consolidation stages in \textbf{b} with the analogous beaker theory. In each sub-schema, synaptic efficacy is defined by the level of the first beaker in relation to the binary threshold (shaded area). The temporal evolution of the first beaker’s liquid level is determined by the hidden state of the second beaker.}
    \label{fig:1}
\end{figure}

Palimpsest consolidation relies on the premise that hidden variables (in the case of biology, resulting from complex biochemical processes) induce changes in synaptic efficacy acting across different timescales \citep{Bhalla2014}. While the operation of such processes is yet to be fully mapped, their phenotypic response can be modelled by the characteristics of fluid diffusion to first approximation \citep{Benna2016}. We can visualise a single synapse as an interconnected chain of progressively larger beakers, where the 1st beaker alone determines the synaptic weight and every subsequent beaker represents a hidden variable. The discrepancies in liquid levels across beakers affect the evolution of liquid distribution over time throughout the entire chain. This is illustrated in Fig. \ref{fig:1}d. While stimulation of the model synapse (adding or removing liquid) occurs exclusively on that first beaker, repeated homopolar stimulation does eventually propagate to the hidden and crucially larger connected beakers farther down the chain. This is the phenomenon of consolidation. However, in a palimpsest memory, competing signals may still successfully write an opposing synaptic state with relatively little effort, albeit in the absence of further reinforcement, the consolidated liquid in the hidden beaker will eventually revert the synapse to expressing the previously consolidated memory.

In this work, we have utilised TiO\textsubscript{2}-based volatile memristive devices (see cross-sections in Fig. \ref{fig:1}a and Methods) which have shown to exhibit bias polarity-dependent bidirectional volatility \citep{Giotis2020}. Volatility is defined as the change in a device’s state within a specified time window and any change in its resistance R that outlasts this window is considered a non-volatile residue. Volatile changes are easily induced but equally short lived (analogous to manipulating the visible synaptic state). Non-volatile residues can accumulate in smaller steps and importantly act as attractors for volatile decay R(t).

Using these characteristics, we first demonstrate palimpsest memory on a single memristive synapse. Specifically, we aim to consolidate a binary state S\textsubscript{1} via LTP and then expressing the antipodal S\textsubscript{0} for a short duration. All plasticity events are uniformly distributed at 30 second intervals (see Methods for details). The results presented in Fig. \ref{fig:1}b-c illustrate 3 distinct consolidation stages. Stage 1: Potentiation events push resistance R to below a predetermined binary threshold R\textsubscript{thres}, corresponding to a short-lived expression of S\textsubscript{1}, but the hidden non-volatile state remains above R\textsubscript{thres}, so in the absence of further reinforcement the synapse reverts to the more consolidated S\textsubscript{0}. Stage 2: Additional potentiation events cause the synapse to undergo LTP. The hidden non-volatile state is pushed below R\textsubscript{thres} and S\textsubscript{1} is consolidated at the long-term timescale. Interestingly, successive potentiation events produce diminishing plastic jumps, observed across stages 1-2, hinting towards the soft resistive state bounds observed in RRAM devices \citep{Messaris2018}. Bounded synaptic efficacy is a known requirement for increasing the capacity of memory networks \citep{Benna2016} and here it results directly from device electrochemistry. Moreover, soft bounding naturally mitigates any asymmetry in device volatility since it allows for increased memory capacity even if potentiation and depression event strengths are not perfectly balanced \citep{Serb2016, Fusi2007}. Stage 3: Competing depression events cause antipodal plastic increases in R, manifesting as a temporary expression of S\textsubscript{0} in memory. However, because the hidden non-volatile state has been consolidated below R\textsubscript{thres}, the observed state overwrite is only realised short-term, reverting to S\textsubscript{1} over time.

Overall, the artificial synapse is able to protect a hidden memory (S\textsubscript{1} caused by LTP), while at the same time reserving the flexibility to express another opposite memory atop it; memory capacity is thus doubled. Next, we note that frequently competing memory events inevitably drive the rigid state closer to R\textsubscript{thres}. The synapse sacrifices stability at the slow timescale as a necessary trade-off for short-term plasticity. Finally, despite our specific device family exhibiting asymmetric responses in potentiation and depression modifications, it is a good candidate for highlighting the consolidation applications due to the high ratio between volatile and non-volatile plasticity changes; a key functional parameter of the system.

\subsection{Operation of memory network}

\begin{figure}[ht]
    \centering
    \includegraphics{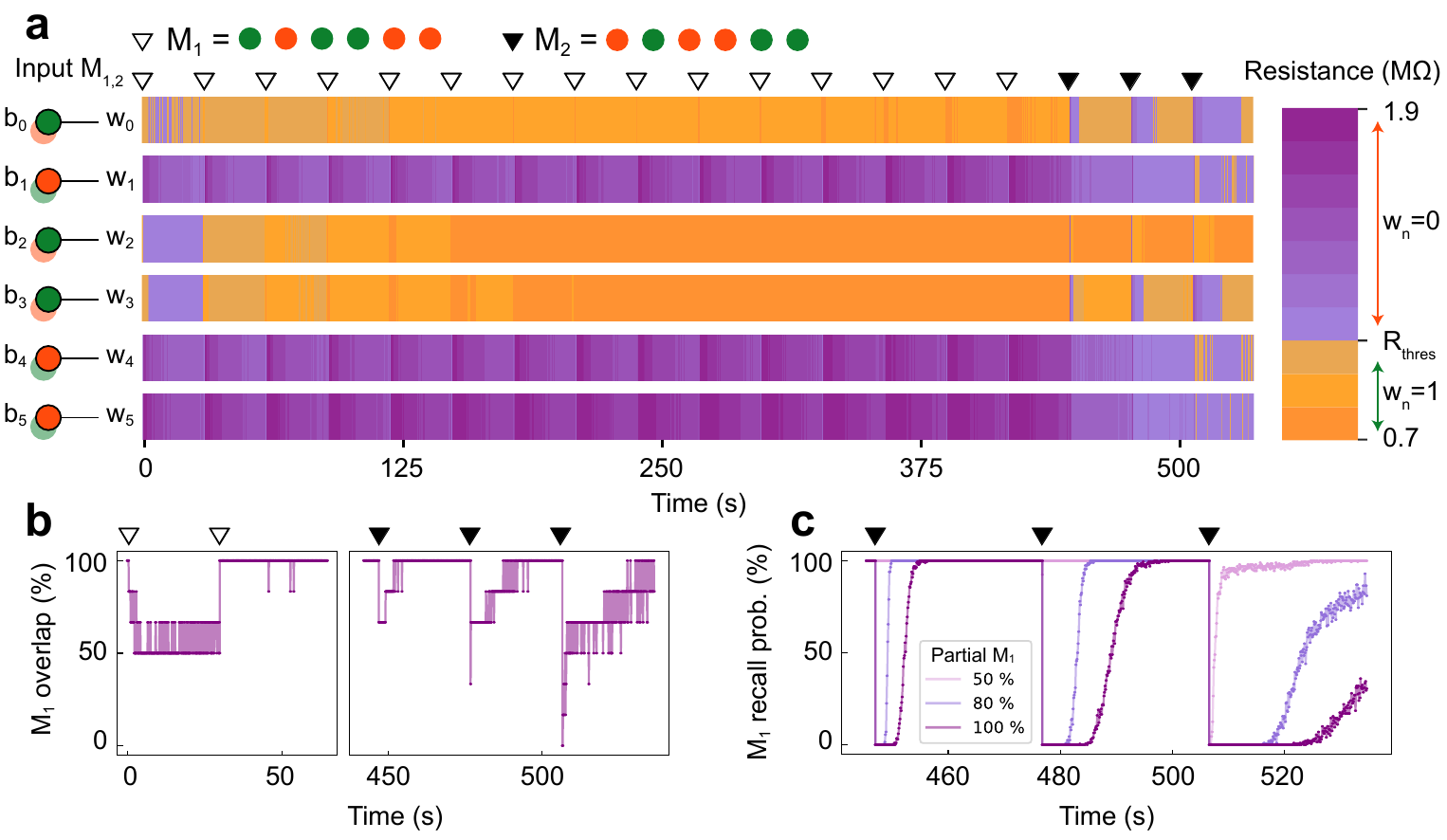}
    \caption{\textbf{Coexistence of short- and long-term memory in memristive synapses.} (a) A set of 6 memristive synapses representing weights w\textsubscript{n} where $n \in \{0,1, ...,5\}$ is fed plasticity instructions b\textsubscript{n} as per patterns M\textsubscript{1} and M\textsubscript{2}. A green dot represents a potentiation instruction and a red dot stands for depression.  Synapses are binary, with weight either 0 for R < R\textsubscript{thres} or 1 otherwise. However, the actual analogue resistance of the memristive synapse allows us to differentiate between ‘shallow’ or ‘deep’ weights. Here, M\textsubscript{1} is stored in the long-term memory using 15 consecutive write events, following which we give the instruction to overwrite M\textsubscript{1} with M\textsubscript{2} 3 consecutive times. (b) The evolution of the memory trace overlap with M\textsubscript{1} under the course of the 18 write events. 100\% overlap translates to perfect storage of M\textsubscript{1} and since M\textsubscript{1} and M\textsubscript{2} are binary and mutually orthogonal, 0\% overlap translates to perfect storage of M\textsubscript{2}. The memory system exhibits the 3 consolidation stages identified in Fig. \ref{fig:1}. At the beginning it takes only three write events to fully consolidate M\textsubscript{1} (at around t = 55s), after which it remains  perfectly  stored in memory until M\textsubscript{2} patterns start arriving at t = 450s. (c) The probability of recalling at least x\% of M\textsubscript{1}, where $x \in \{50\%,80\%,100\%\}$. Traces are averages from 200 simulated experimental runs, each under the addition of noise (see Methods section). Recovering only part of M\textsubscript{1} becomes possible earlier as the desired degree of reconstruction drops.}
    \label{fig:2}
\end{figure}

Next, we construct a small memristive network comprised of 6 synapses to consolidate two competing signals, M\textsubscript{1} ($\medtriangledown$ = 101100) and M\textsubscript{2} ($\filledmedtriangledown$ = 010011). These are antipodal binary vectors and consequently, competition for storage consists a worst-case zero-sum game. The experimental setup is explained in Fig. \ref{fig:2}a. Each bit b\textsubscript{n}, $n \in \{0,1,...,5\}$, of memories M\textsubscript{1,2} is written on a corresponding memristive synapse w\textsubscript{n} (see Methods). Encoding palimpsest memories in this worst-case scenario, implies that system performance would only increase in generalised applications: for instance, random uncorrelated binary vectors average a 50\% similarity rate leaving only the 50\% subject to destructive interference effects.

Examining the analogue resistance values gives further insight on how the two memory signals interact across the synapses. Individual synapses cycle through the consolidation stages outlined in Fig. \ref{fig:1}. Progressive modifications push the hidden state further away from R\textsubscript{thres} as reflected from the ‘deeper’ resistance values, at which point, M\textsubscript{1} is strongly consolidated at the long-term. When M\textsubscript{2} is presented in memory, individual synapses are resistant to encoding the requested bit states. Specifically, synapses w\textsubscript{1,4,5} that have undergone LTD fail to fully encode the respective M\textsubscript{2} states in the two first write events. This stems from the mentioned asymmetry between the devices’ volatile responses in opposite directions. However, all synapses are pushed closer to R\textsubscript{thres} suggesting a retreat from strong M\textsubscript{1} consolidation. In the final M\textsubscript{2} event, the new memory is fully written at the short-term, before M\textsubscript{1} is reinstated.

Memory performance is macroscopically examined in Fig. \ref{fig:2}b. The overlap between M\textsubscript{1} and the system’s state is shown against time. Due to the antipodal relationship between M\textsubscript{1} and M\textsubscript{2}, an x\% overlap between the system and the former implies a (100-x)\% overlap with the latter. The time axis has been truncated since everywhere between the second writing of M\textsubscript{1} at t = 50s and the first presentation of M\textsubscript{2} at t = 450s the overlap with M\textsubscript{1} is solidly 100\%. This is also evident by examining Fig. \ref{fig:2}a. The signal overlap is visibly quantised due to the small size of the synaptic circuit. Ultimately, each presentation of M\textsubscript{2} is progressively more successful at overwriting the consolidated M\textsubscript{1}, whose recovery becomes progressively slower yet still achievable. Interestingly, reinstation of M\textsubscript{1} is non-monotonic. This is attributable to noise, which becomes a deciding factor when R is close to R\textsubscript{thres}.

To further quantify how noise is expected to affect memory performance we run the following experiment: First, using existing volatility modelling methods\citep{Giotis2020b} we obtain an expected ideal estimate of behaviour under the stimulation protocol of Fig. \ref{fig:2}a. Next, calculate the noise distribution observed in real device data. Finally, run 200 simulations, each contaminated with different random noise and evaluate recall performance of M\textsubscript{1} after the presentations of M\textsubscript{2} at the end of each run (see Methods for details). Fig. \ref{fig:2}c shows the probability of recalling M\textsubscript{1} fully (100\% signal overlap) or partially (at least 80\% or 50\% overlap) after applying the M\textsubscript{2} write events. As expected, the ability for recall of M\textsubscript{1} drops after each presentation of M\textsubscript{2}, translating into longer recovery times and possibly lower achievable maximum recall values. Nevertheless, by the 3rd M\textsubscript{2} presentation the system continues to achieve at least 80\% recall more than 80\% of the time. This illustrates that the degradation of recall performance as competing memories progressively overwrite each other is smooth, which implies that effective performance will be affected by the required recall accuracy. Most AI systems already function on the basis of average estimations of learnt signals, which implies a good quality partial recall suffices in practice\citep{LeCun2015}.

\begin{figure}[ht]
    \centering
    \includegraphics{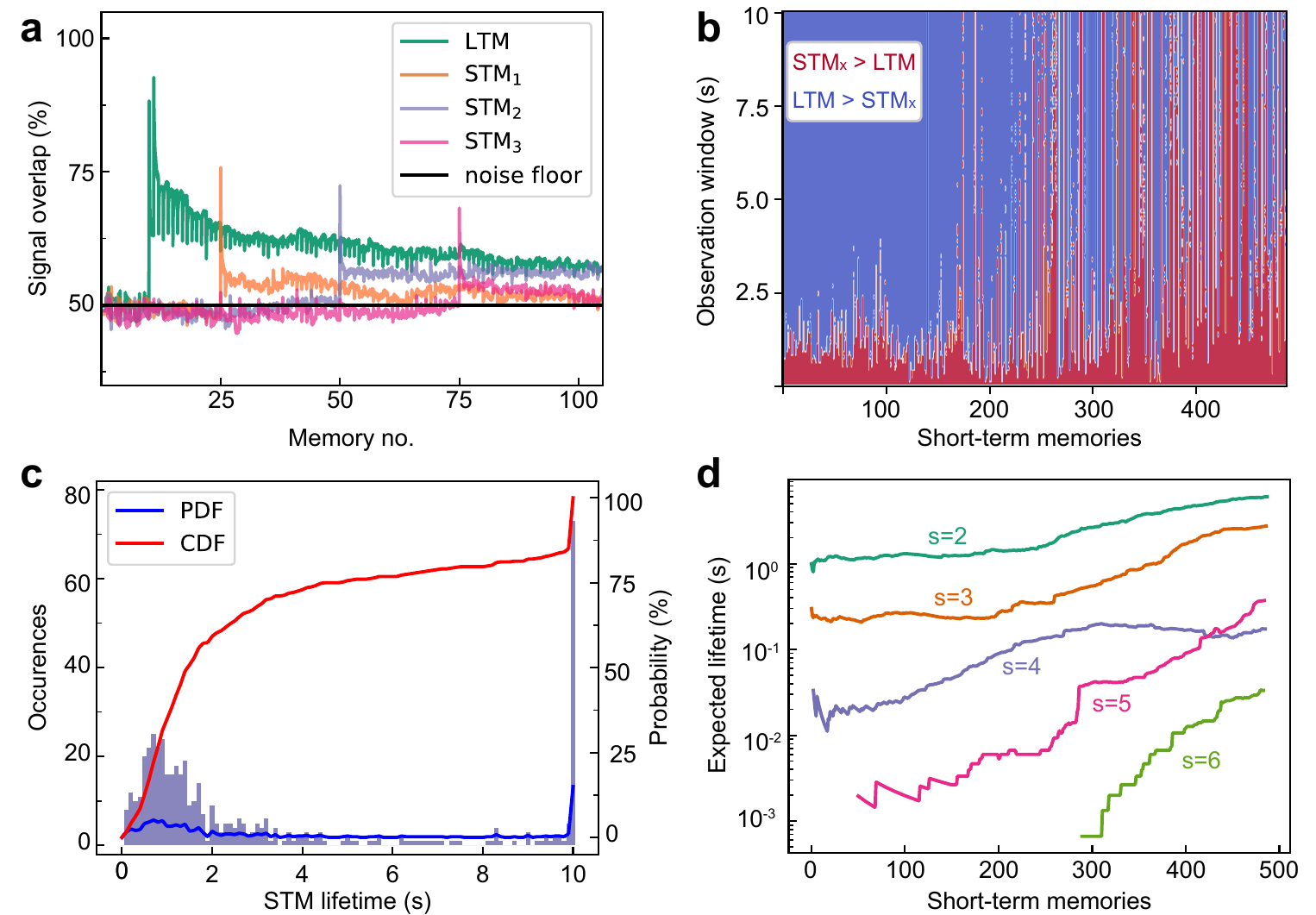}
    \caption{\textbf{Memory performance under continuous stimulation.} (\textbf{a}) Evolution of a long-term memory (LTM) and following short-term memory (STM) signals. Each randomly chosen STM, briefly surpasses LTM before the latter reinstates its dominance. Importantly, decaying signals retain higher than chance overlaps (they remain above the 50\% noise floor). (\textbf{b}) Observation of all random STMs after LTM consolidation. Red regions indicate the total time STM\textsubscript{x} signal is stronger than LTM; blue regions suggest the opposite. (\textbf{c}) Histogram of distribution of all STM total lifetimes on the left y-axis. The corresponding PDF and CDF curves are illustrated via the right y-axis. (\textbf{d}) Weighted expected STM lifetime for different LTM consolidation strengths (s). The green s = 2 line corresponds to the memory data presented in \textbf{a-c}.}
    \label{fig:3}
\end{figure}

The performance of the memory can be generalised even further by considering palimpsest behaviour within the easier context of random and uncorrelated memory traces. In these runs, emphasis has been shifted from utilising very large plastic capabilities that magnify the contrast between short-term and long-term memory timescales (see Fig. \ref{fig:1} \ref{fig:2}) to ensuring symmetrical performance under potentiation and depression modifications (see Methods). In practice, we have sacrificed the high ratio between volatile jump and non-volatile residues for symmetrical, bidirectional volatile responses.

A memory network comprised of 100 synapses is subject to an ongoing stream of 500 input memories that are chosen randomly. Synaptic modifications occur evenly spaced every 10 seconds. In Fig. \ref{fig:3}a-c, 10 random memories are written in the system before a long-term memory (LTM) is consolidated with an intensity of s = 2 repetitions. A relatively low value for s has been chosen on purpose, in order to prevent a very deep entrenchment of LTM in the rigid timescale. As observed in Fig. \ref{fig:3}a, this prevents LTM from being fully written in the system. The memory overlap with 3 randomly chosen input memories (short-term memories – STMs) is also shown against the 50\% noise floor. Spikes mark the presentation of these memories to the system. The general LTM overlap is surpassed by STM\textsubscript{1-3} immediately after these are written in the system, before LTM is reinstated as the strongest memory. Some significant observations deserve mention here. First, LTM’s failure to achieve a perfect signal overlap can be explained by the preceding plasticity events. Stochastic stimulation can cause spontaneous entrenchment of random synapses which then challenges the consolidation of LTM. Second, the overall overlap of both LTM and successive STM signals falls over time, highlighting the slow but continuous degradation of memories as new inputs are received. Lastly, all STM signals retain an above-chance representation strength for a long time even if the interval during which they are the dominant memory is short. Hence, even at one-shot scenarios, the network exhibits high capacity in familiarity recalls, meaning it can distinguish whether multiple memories have been presented before or not.

Because acceptance thresholds for absolute signal overlaps are directly related to specific application needs, we focus on the relative strength difference between the consolidated LTM and incoming STMs. Fig. \ref{fig:3}b shows which memory is dominant (has the highest degree of correlation with the actual state of the memory network) in 10s observation windows following each presentation of an STM. The LTM has been consolidated just before the commencement of the first STM trial. Blue regions indicate LTM dominance and red regions indicate STM dominance. For the first 100 random STMs the 10s observation window is dominated by (quickly restored) LTM signals. However, as more patterns are presented to the memory, the consolidated LTM pattern degrades, and more recent signals prevail; red sections become longer and denser as LTM restoration collapses.

STM lifetime is defined as the total time period where some STMx signal dominates over the LTM (see Fig. \ref{fig:3}c). The histogram of lifetime occurrences is shown on the left y-axis, while the corresponding probability and cumulative distributions are shown on the right y-axis. The data can be split into 3 main segments: First, a large bulk of STMs surviving between 0 and 2 seconds, mainly populated by STMs presented towards the beginning of the test and representing about 50\% of the signals. Second, a relatively sparsely populated trough between ~2 and 10 seconds reflects the fact that after ~2s the volatile component of the synaptic dynamics has for the most part relaxed. It should be noted that there is some preliminary evidence to suggest that volatility might function on phases; characterised by an initial phase of rapid decay, followed by a slower decay of the residue at a much smaller time constant \citep{Giotis2020}. This subtle phenomenon, however, requires further study. Finally, the peak at 10s bins together any cases where the LTM would either be restored at more than 10s or fail to be restored and thus appears as a prominent peak. Overall, while LTM remains consolidated in the rigid timescale, the network tends to rebound to it within that 2s timeframe. This is highlighted by the lifetime cumulative distribution (CDF) shown in red. The probability that some STM\textsubscript{x} survives for up to 2s is approximately 60\%, while the probability that it never gets written is about 20\%.

We also note that the system is able to retain long-term memories more rigidly if they are more intensely entrenched/consolidated. The same experiment has been repeated for a range of LTM consolidation intensities s = \{2,3,4,5,6\}. In Fig. \ref{fig:3}d we show averaged STM lifetimes over a sliding window of 150 STM presentations for each consolidation intensity (data points before STM no. 150 only average over all previous STMs). As consolidation intensity increases, the ability to write any STM on top of LTM reduces significantly. Early lifetimes decrease by about 1 order of magnitude per s level until the first few tens (s=5) or hundreds (s=6) of STMs fail to surpass the LTM completely (expected lifetime=0s). This is because increased consolidation of LTM causes the rigid non-volatile residues to shift further away from the (binary) efficacy threshold, reducing the efficiency of potentiation/depression events in changing the weight. Metaplastic properties are thus realised implicitly by our synapses. This shows how the trade-off between capacity and recollection accuracy can be controlled: high levels of consolidation appear to safeguard the LTM against at least hundreds of incoming STMs.

\begin{figure}[ht]
    \centering
    \includegraphics{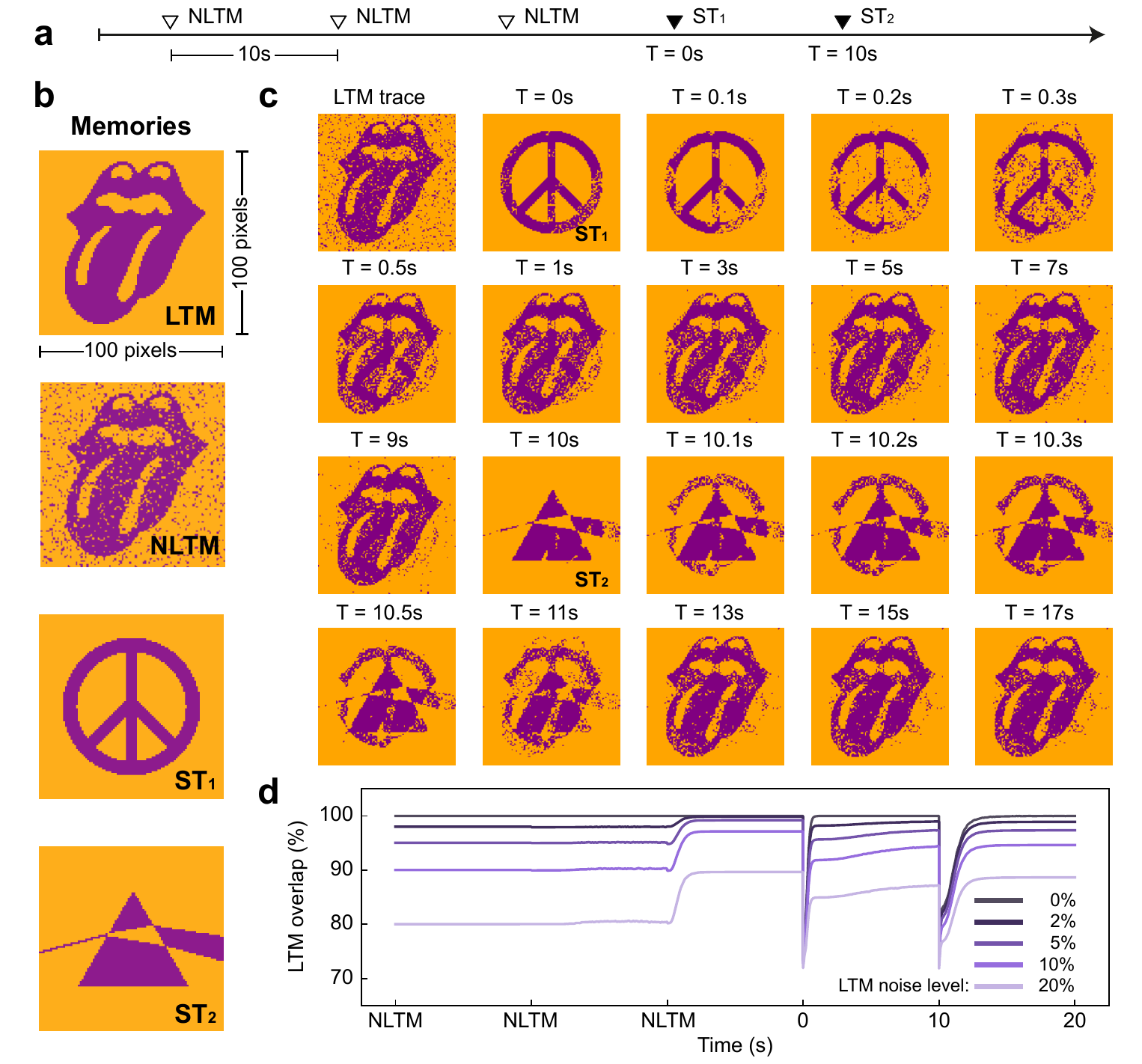}
    \caption{\textbf{Consolidation of binary images.} (\textbf{a}) Operation timeline - images are written in memory every 10 seconds. Random noisy representations of a long-term memory (NLTM and LTM respectively) are written 3 times before two images are written, once each, in the short-term (ST\textsubscript{1,2}). (\textbf{b}) Original signals of image memories. An NLTM depiction is chosen at random. (\textbf{c}) Temporal snapshots of memory states. The initial snapshot corresponds to last observation before ST\textsubscript{1} is seen. Time is referenced after that instant. The moments when the two ST signals are written are noted. In both cases, a noisy trace of each ST signal can be accessed in the memory immediately after observation. Then, the ST signal gradually degrades while LTM is reinstated in memory. (\textbf{d}) Outline of the LTM signal overlap, as it evolves following the timeline in \textbf{a}. The initial overlap reflects the corresponding noise level. The experiment is repeated for different NLTM noise levels. Even if a noiseless version of LTM is never presented to the system per se, the memory achieves above-baseline overlaps vis-a-vis LTM after consolidation has finished. Importantly, this high overlap is retained even after the ST\textsubscript{1} and ST\textsubscript{2} are seen by the network.}
    \label{fig:4}
\end{figure}

\subsection{Unsupervised memory reconstruction}

After evaluating performance with random memory streams, we examined the network’s operation with statistically correlated signals. We are interested in whether the system can identify statistical significance without supervision. To address this, we have set up a synaptic network comprised of 100 artificial synapses that observes incoming binary images (see Methods). We consolidate one image in the long-term memory (LTM), then sequentially store two images ST\textsubscript{1,2} in short-term capacity. Additionally, we test the network’s ability to consolidate only the relevant information within LTM. To do this, we present the network with 3 independently noise-contaminated LTM variations (NLTM), before writing ST\textsubscript{1} and ST\textsubscript{2}. Thus, signal correlation is observed at two levels: All different images are at least 70\% similar to each other while the noisy NLTM variations are at least 80\% similar. The timeline of synaptic modifications is shown in Fig. \ref{fig:4}a. Time is set to T = 0s at the occurrence of ST\textsubscript{1}.

The image memories are shown in Fig. \ref{fig:4}b. Each pixel corresponds to a memristive synapse. A random NLTM is illustrated. Snapshots of the network’s evolution are shown in Fig. \ref{fig:4}c. In this scenario, NLTMs are contaminated at a 10\% noise level. The first snapshot (top left corner) shows the LTM trace just before ST\textsubscript{1} is written in the network. At T = 0s, ST\textsubscript{1} successfully overtakes the network for a short timeframe. Later snapshots show the network’s transition to its long-term state. By time T = 9s the network has recovered LTM. Importantly, the system’s state at that point constructs a visibly cleaner representation of LTM than the final NLTM trace shown before ST\textsubscript{1} is presented. At T = 10s, the third image, ST\textsubscript{2}, is written in the memory network. Noticeably, the decay of ST\textsubscript{2} is first caused by the pixels that are common in LTM and ST\textsubscript{1}. These have been consolidated by both memories and are thus reinstated faster. Eventually, LTM becomes dominant again at time T = 17s.

By comparing the LTM trace in the top left corner and the last snapshot, we observe that the network is able to automatically denoise consolidated signals. This property arises naturally from the fact that random information, presented only sparsely, is less susceptible to consolidation than bits written with higher intensity, as shown in Fig. \ref{fig:3}. Hence, the network is able to average over many noisy representations and converge closer to the actual, but never explicitly presented, LTM signal. These denoising properties are clearly visible in Fig. \ref{fig:4}d, where the experiment has been repeated for several noise levels on LTM and average signal overlaps are shown versus time. The overlap at the time of the first NLTM occurrence reflects the corresponding noise levels. However, after writing all NLTM signals, the overlap between the network and the actual LTM signal is increased significantly entirely spontaneously; by almost 10\% at the limit scenario of 20\% noise level. Importantly, the representation of LTM remains at above initial noise levels even after the modifications induced by ST\textsubscript{1,2}.

\section{Discussion}

In this work we focused on binary synapses, which are known to support adequate learning in mathematical models \citep{Benna2017} as well as deep-learning algorithms \citep{Laborieux2021}. The weight of the synapse is a binarized version of its resistive state and the interplay between intense bidirectional volatility and small non-volatile residues underlies its palimpsest capability. We note that unidirectional volatility is already sufficient to support consolidation, i.e. the transition from short- to long-term memory \citep{Chang2011, Ohno2011}. This concept can be naturally extended to synaptic weights of higher resolution.

Another remark about this palimpsest memory is that the contents of the memory are in general imperfect reflections of the desired memory. This is not unusual per se since neuro-inspired systems work on the basis of imperfect information typically by default (classifiers sort noisy inputs into neat classes), but in palimpsest memories we have the additional factor of LTM-STM relations to consider. Finally, we note that this technology bears some interesting similarities to how real estate is used for multiple storage in visual working memory systems \citep{Matthey2015, Luck1997, Zhang2008}. This partition of memory storage is an advantageous adaptation since only information that is relevant to a specific cognitive task is needed for undergoing the said task. The dual capacity that is exhibited by our devices resembles the bistable switching that is known to govern synaptic plasticity \citep{Bhalla2014}. Specifically, the accumulation of non-volatile residues after LTP/LTD can be thought of as an equivalent mechanism to calcium/calmodulin-dependent protein kinase II (CaMKII), which is considered to be a primary molecular memory mechanism \citep{Lisman2002, Miller2005}.

In conclusion, we have demonstrated how memristive synapses can underpin the full functionality of a palimpsest memory without using specialised consolidation signals or other complex signalling. The physics of the device thus not only allow metaplasticity (state-dependent plasticity) and consolidation (progressive entrenchment of memories), but also the expression of (multiple) short-term memories atop a consolidated, long-term one. This is achieved even with complete anti-correlated memories, which is a feature uniquely defining palimpsest capability. This flexibility and LTM reconstructive ability can enhance the performance of in-memory computing \citep{Boybat2021, LeGallo2018, Hu2015} where systems are required to adapt quickly to incoming stimuli and is of direct relevance to neuro-inspired computing. Moreover, the short-lived span of overlapping memories resembles short-term attention mechanisms, which have recently shown promise towards more complex AI algorithms \citep{Vaswani2017}. Attention mechanisms can also be implemented using the high-capacity STM familiarity filters that are exhibited here (a familiarity filter is a memory that recognises when a memory input is present inside the memory even if it no longer possesses enough information to reconstruct the memory). Our memory also implements unsupervised (LTM) memory reconstruction in hardware, supporting previously linked theories of consolidation \citep{Benna2016, Fusi2005} and optimal recall in the CA3 area of the hippocampus \citep{Savin2014}. We thus anticipate that this work represents a significant milestone towards realising a wide range of neuro-inspired systems practically, in silico.

\section{Methods}\label{methods}

\textbf{Device fabrication.} Devices used in this work are vertical Metal-Insulator-Metal (MIM) structures with electrode dimensions of 20x20\textmu m\textsuperscript{2} as depicted in Fig. \ref{fig:1}a.  The initial fabrication step was to thermally grow 200nm of SiO\textsubscript{2} on top of 6-inch Silicon wafers, which were used as substrates for the process. This thermal oxide serves as an insulator, separating all devices from the Silicon substrate. Each of the three layers in the MIM structure was deposited by following a four-step process, namely, Lithography, Short reactive ion etching, Deposition and Lift-off. Lithography was completed with an EVG 620 TB mask aligner to expose each mask pattern on a negative tone AZ-2070 resist. After each lithography step, a short O\textsubscript{2} plasma cleaning step ensures cleanliness of the area which has been prepared for material deposition, by removing resist residuals. The deposition step was completed by E-beam evaporation for metal materials and by magnetron sputtering for the active layer material. Bottom electrodes were deposited using the Leybold LAB 700 E-beam equipment. Initially 5nm of Titanium (Ti) adhesion layer was deposited on top of the thermal oxide and, in continuation, 20nm of Gold (Au) was deposited. After bottom layer deposition the wafer was soaked in NMP overnight for lift-off. The middle layer consists of TiO\textsubscript{2} deposited with an EvoVac angstrom engineering DC sputtering equipment. The active layer consists of 25nm of TiO\textsubscript{2}; sputtered at room temperature from a metal Ti target in a 4\% O\textsubscript{2}/Ar atmosphere and 3mTorr pressure at 200W. Following the active layer deposition step, lift-off was carried out with the Optiwet-ST30 tool, which ensures a clean lift-off by spraying hot NMP (60 Celsius) with 3mbar pressure on the wafer for 30 minutes. Lastly, top electrodes were deposited by following the same process as described for the bottom electrodes, with the deposited material in this case being 12nm of Platinum (Pt).

\textbf{Memristive synapse set-up.} For all of our experiments, single volatile devices were operated in a binary fashion, dictated by resistance R compared to a chosen threshold value R\textsubscript{thres}. Plasticity changes were induced following the rule in Eq. \ref{eq:1}. Binary signals equal to ‘1’ induce potentiation events while signals equal to ‘0’ are causing the synapses to become depressed.

\begin{equation}\label{eq:1}
    plasticity\ event =
    \begin{cases}
      potentiation, & \text{if}\ V>1 \\
      depression, & \text{otherwise}
    \end{cases}
\end{equation}

Accordingly, the binary weight w was computed using Eq. \ref{eq:2}.

\begin{equation}\label{eq:2}
    w =
    \begin{cases}
      1, & \text{if}\ R < R_{thres} \\
      0, & \text{otherwise}
    \end{cases}
\end{equation}

\textbf{RRAM volatility modelling and noise extraction.} Memristive volatility was quantified using existing modelling methods \citep{Giotis2020b}. Specifically, R(t) was expressed via Eq. \ref{eq:3}.

\begin{equation}\label{eq:3}
    R(t) = \alpha e^{\big(\frac{t}{\tau} \big)^{\beta}} + \gamma
\end{equation}

Volatility noise was calculated as the percentage (\%) difference between the ideal model and real device as shown in Eq. \ref{eq:4}. RRAM noise was seen to follow a characteristic normal distribution. By extracting the distribution’s mean value \textmu and standard deviation \textsigma, noise could then be added stochastically on ideal data using Eq. \ref{eq:5}.

\textbf{Consolidation of fully destructive memories.} To begin our study, we wished to examine our networks performance on the worst-case scenario of two antipodal and fully competing binary memories. We used 6 memristive synapses that were independently stimulated such that M\textsubscript{1} = [101100] and M\textsubscript{2} = [010011] were written in the long- and short-term timescales respectively. Specifically, M\textsubscript{1} and M\textsubscript{2} were presented in memory for 15 and 3 consecutive times respectively. For this conceptual demonstration of our technology, we biased our devices using single \textpm7V stimuli for a total duration of 100μs, while analogue states were read at 0.5V. This profile yielded a significant contrast ratio between plastic and rigid efficacy changes which has allowed a clearer depiction of palimpsest state overwrites. Plasticity changes where induced following the rule in Eq. \ref{eq:1} and the corresponding synaptic states or ‘weights’ were calculated using Eq. \ref{eq:2}. A binary threshold value R\textsubscript{thres} was explicitly chosen such that individual device histories achieve the best overlap possible.

\begin{equation}\label{eq:4}
    (\%)\Delta R = noise = \frac{R_{ideal} - R}{R} \times 100\%
\end{equation}

\begin{equation}\label{eq:5}
    p(noise) = \frac{1}{\sqrt{2\pi \sigma ^2}} e^{\frac{(noise-\mu)^2}{2\sigma ^2}}
\end{equation}

\textbf{Random memory stream.} For this study, we aimed at evaluating our technology’s ability to consolidate memories that were uncorrelated in nature. Moreover, we required a network sufficiently large to reflect the statistics of its performance in a smooth manner and avoid the quantisation errors shown in Fig. \ref{fig:2}. Hence, we devised a network comprised of 100 identical memristive synapses by utilising Eq. 3-5. The synapses’ binary threshold was extracted via applying alternating plasticity events and observing the natural occurring equilibrium position. In order to evaluate the technology’s performance at a memory level, our main priority has been the symmetrical response to LTP/LTD events, such that no binary state is consolidated de facto over time. Devices have this time been stimulated using 500 train pulses (500\textmu s width each) at 1.4V and -2.6V for potentiation and depression events respectively. Resistance was read at 1.0V. This profile utilises asymmetric stimulation energy but ensures equal writing speeds. Our choices have been made while ignoring the energy efficiency of our systems in favour of conceptual and operational clarity. However, bidirectional volatility has been reported in HfO\textsubscript{2} based memristors with programming voltages as low as 0.3V \citep{Covi2019}, which is a promising pathway towards less energy consuming solutions.

\textbf{Consolidation of binary images.} In this section, we utilised a memory network of 100x100 identical synapses to consolidate binary images. These are the same synapses that were described in the above Methods section and the same operation profile was employed. The first image was implicitly reconstructed in the long-term memory using noisy variation of the original signal. Noise was added to the signal by independently choosing to flip each bit with a probability p = {0\%, 2\%, 5\%, 10\%, 20\%}. 

\bibliographystyle{unsrtnat}
\bibliography{references}  %%% Uncomment this line and comment out the ``thebibliography'' section below to use the external .bib file (using bibtex) .

\begin{thebibliography}{41}
\providecommand{\natexlab}[1]{#1}
\providecommand{\url}[1]{\texttt{#1}}
\expandafter\ifx\csname urlstyle\endcsname\relax
  \providecommand{\doi}[1]{doi: #1}\else
  \providecommand{\doi}{doi: \begingroup \urlstyle{rm}\Url}\fi

\bibitem[Shepherd(2004)]{Shepherd2004}
Gordon~M. Shepherd.
\newblock \emph{{The Synaptic Organization of the Brain}}.
\newblock Oxford University Press, jan 2004.
\newblock ISBN 9780195159561.
\newblock \doi{10.1093/acprof:oso/9780195159561.001.1}.
\newblock URL
  \url{https://oxford.universitypressscholarship.com/view/10.1093/acprof:oso/9780195159561.001.1/acprof-9780195159561}.

\bibitem[Koch(1998)]{Koch1998}
Christof Koch.
\newblock \emph{{Biophysics of Computation: Information Processing in Single
  Neurons}}.
\newblock Oxford University Press, nov 1998.
\newblock ISBN 9780195104912.
\newblock \doi{10.1093/oso/9780195104912.001.0001}.
\newblock URL
  \url{https://oxford.universitypressscholarship.com/view/10.1093/oso/9780195104912.001.0001/isbn-9780195104912}.

\bibitem[Lecun et~al.(2015)Lecun, Bengio, and Hinton]{LeCun2015}
Yann Lecun, Yoshua Bengio, and Geoffrey Hinton.
\newblock \emph{{Deep learning}}, volume 521.
\newblock Nature Publishing Group, may 2015.
\newblock \doi{10.1038/nature14539}.
\newblock URL \url{http://www.nature.com/articles/nature14539}.

\bibitem[Xu et~al.(2018)Xu, Ding, Hu, Niemier, Cong, Hu, and Shi]{Xu2018}
Xiaowei Xu, Yukun Ding, Sharon~Xiaobo Hu, Michael Niemier, Jason Cong, Yu~Hu,
  and Yiyu Shi.
\newblock {Scaling for edge inference of deep neural networks}.
\newblock \emph{Nature Electronics}, 1\penalty0 (4):\penalty0 216--222, apr
  2018.
\newblock ISSN 2520-1131.
\newblock \doi{10.1038/s41928-018-0059-3}.
\newblock URL \url{https://doi.org/10.1038/s41928-018-0059-3
  http://www.nature.com/articles/s41928-018-0059-3}.

\bibitem[Benna and Fusi(2016)]{Benna2016}
M~K Benna and S~Fusi.
\newblock {Computational principles of synaptic memory consolidation}.
\newblock \emph{Nat Neurosci}, 19\penalty0 (12):\penalty0 1697--1706, 2016.
\newblock \doi{10.1038/nn.4401}.
\newblock URL \url{https://www.ncbi.nlm.nih.gov/pubmed/27694992}.

\bibitem[Abraham(2008)]{Abraham2008}
Wickliffe~C Abraham.
\newblock {Metaplasticity: Tuning synapses and networks for plasticity}.
\newblock \emph{Nature Reviews Neuroscience}, 9\penalty0 (5):\penalty0
  387--399, 2008.
\newblock ISSN 1471003X.
\newblock \doi{10.1038/nrn2356}.
\newblock URL \url{www.nature.com/reviews/neuro}.

\bibitem[Stathopoulos et~al.(2017)Stathopoulos, Khiat, Trapatseli, Cortese,
  Serb, Valov, and Prodromakis]{Stathopoulos2017}
S~Stathopoulos, A~Khiat, M~Trapatseli, S~Cortese, A~Serb, I~Valov, and
  T~Prodromakis.
\newblock {Multibit memory operation of metal-oxide bi-layer memristors}.
\newblock \emph{Sci Rep}, 7\penalty0 (1):\penalty0 17532, 2017.
\newblock \doi{10.1038/s41598-017-17785-1}.
\newblock URL \url{https://www.ncbi.nlm.nih.gov/pubmed/29235524}.

\bibitem[Serb et~al.(2016)Serb, Bill, Khiat, Berdan, Legenstein, and
  Prodromakis]{Serb2016}
A~Serb, J~Bill, A~Khiat, R~Berdan, R~Legenstein, and T~Prodromakis.
\newblock {Unsupervised learning in probabilistic neural networks with
  multi-state metal-oxide memristive synapses}.
\newblock \emph{Nat Commun}, 7:\penalty0 12611, 2016.
\newblock \doi{10.1038/ncomms12611}.
\newblock URL \url{https://www.ncbi.nlm.nih.gov/pubmed/27681181}.

\bibitem[Gupta et~al.(2016)Gupta, Serb, Khiat, Zeitler, Vassanelli, and
  Prodromakis]{Gupta2016}
Isha Gupta, Alexantrou Serb, Ali Khiat, Ralf Zeitler, Stefano Vassanelli, and
  Themistoklis Prodromakis.
\newblock {Real-time encoding and compression of neuronal spikes by metal-oxide
  memristors}.
\newblock \emph{Nature Communications}, 7\penalty0 (1):\penalty0 12805, dec
  2016.
\newblock ISSN 20411723.
\newblock \doi{10.1038/ncomms12805}.
\newblock URL \url{http://www.nature.com/articles/ncomms12805}.

\bibitem[Berdan et~al.(2016)Berdan, Vasilaki, Khiat, Indiveri, Serb, and
  Prodromakis]{Berdan2016}
R~Berdan, E~Vasilaki, A~Khiat, G~Indiveri, A~Serb, and T~Prodromakis.
\newblock {Emulating short-term synaptic dynamics with memristive devices}.
\newblock \emph{Sci Rep}, 6:\penalty0 18639, 2016.
\newblock \doi{10.1038/srep18639}.
\newblock URL \url{https://www.ncbi.nlm.nih.gov/pubmed/26725838}.

\bibitem[Yoon et~al.(2018)Yoon, Wang, Kim, Wu, Ravichandran, Xia, Hwang, and
  Yang]{Yoon2018}
Jung~Ho Yoon, Zhongrui Wang, Kyung~Min Kim, Huaqiang Wu, Vignesh Ravichandran,
  Qiangfei Xia, Cheol~Seong Hwang, and J.~Joshua Yang.
\newblock {An artificial nociceptor based on a diffusive memristor}.
\newblock \emph{Nature Communications}, 9\penalty0 (1), 2018.
\newblock ISSN 20411723.
\newblock \doi{10.1038/s41467-017-02572-3}.
\newblock URL \url{www.nature.com/naturecommunications}.

\bibitem[Boybat et~al.(2018)Boybat, {Le Gallo}, Nandakumar, Moraitis, Parnell,
  Tuma, Rajendran, Leblebici, Sebastian, and Eleftheriou]{Boybat2021}
Irem Boybat, Manuel {Le Gallo}, S.~R. Nandakumar, Timoleon Moraitis, Thomas
  Parnell, Tomas Tuma, Bipin Rajendran, Yusuf Leblebici, Abu Sebastian, and
  Evangelos Eleftheriou.
\newblock {Neuromorphic computing with multi-memristive synapses}.
\newblock \emph{Nature Communications}, 9\penalty0 (1):\penalty0 1--12, dec
  2018.
\newblock ISSN 20411723.
\newblock \doi{10.1038/s41467-018-04933-y}.
\newblock URL \url{www.nature.com/naturecommunications}.

\bibitem[Abraham and Bear(1996)]{Abraham1996}
Wickliffe~C. Abraham and Mark~F. Bear.
\newblock {Metaplasticity: The plasticity of synaptic plasticity}.
\newblock \emph{Trends in Neurosciences}, 19\penalty0 (4):\penalty0 126--130,
  apr 1996.
\newblock ISSN 01662236.
\newblock \doi{10.1016/S0166-2236(96)80018-X}.
\newblock URL
  \url{https://www.sciencedirect.com/science/article/pii/S016622369680018X?via%3Dihub}.

\bibitem[Fusi et~al.(2005)Fusi, Drew, and Abbott]{Fusi2005}
S~Fusi, P~J Drew, and L~F Abbott.
\newblock {Cascade models of synaptically stored memories}.
\newblock \emph{Neuron}, 45\penalty0 (4):\penalty0 599--611, 2005.
\newblock \doi{10.1016/j.neuron.2005.02.001}.
\newblock URL \url{https://www.ncbi.nlm.nih.gov/pubmed/15721245}.

\bibitem[Brivio et~al.(2019)Brivio, Conti, Nair, Frascaroli, Covi, Ricciardi,
  Indiveri, and Spiga]{Brivio2019}
S~Brivio, D~Conti, M~V Nair, J~Frascaroli, E~Covi, C~Ricciardi, G~Indiveri, and
  S~Spiga.
\newblock {Extended memory lifetime in spiking neural networks employing
  memristive synapses with nonlinear conductance dynamics}.
\newblock \emph{Nanotechnology}, 30\penalty0 (1), 2019.
\newblock ISSN 13616528.
\newblock \doi{10.1088/1361-6528/aae81c}.
\newblock URL \url{https://doi.org/10.1088/1361-6528/aae81c}.

\bibitem[Demirag et~al.(2021)Demirag, Moro, Dalgaty, Navarro, Frenkel,
  Indiveri, Vianello, and Payvand]{Demirag2021}
Yigit Demirag, Filippo Moro, Thomas Dalgaty, Gabriele Navarro, Charlotte
  Frenkel, Giacomo Indiveri, Elisa Vianello, and Melika Payvand.
\newblock {PCM-trace: Scalable synaptic eligibility traces with resistivity
  drift of phase-change materials}.
\newblock \emph{Proceedings - IEEE International Symposium on Circuits and
  Systems}, 2021-May:\penalty0 1--6, 2021.
\newblock ISSN 02714310.
\newblock \doi{10.1109/ISCAS51556.2021.9401446}.

\bibitem[Zhu et~al.(2017)Zhu, Du, Jeong, and Lu]{Zhu2017}
Xiaojian Zhu, Chao Du, Yeonjoo Jeong, and Wei~D Lu.
\newblock {Emulation of synaptic metaplasticity in memristors}.
\newblock \emph{Nanoscale}, 9\penalty0 (1):\penalty0 45--51, 2017.
\newblock ISSN 20403372.
\newblock \doi{10.1039/c6nr08024c}.
\newblock URL \url{www.rsc.org/nanoscale}.

\bibitem[Wu et~al.(2018)Wu, Wang, Luo, Banerjee, Cao, Zhang, Wu, Liu, Li, and
  Liu]{Wu2018}
Quantan Wu, Hong Wang, Qing Luo, Writam Banerjee, Jingchen Cao, Xumeng Zhang,
  Facai Wu, Qi~Liu, Ling Li, and Ming Liu.
\newblock {Full imitation of synaptic metaplasticity based on memristor
  devices}.
\newblock \emph{Nanoscale}, 10\penalty0 (13):\penalty0 5875--5881, apr 2018.
\newblock ISSN 20403372.
\newblock \doi{10.1039/c8nr00222c}.

\bibitem[Cheng et~al.(2018)Cheng, Li, Zhang, Fang, Zhu, Liu, Xu, Cai, Yan,
  Yang, and Huang]{Cheng2018}
Caidie Cheng, Yiqing Li, Teng Zhang, Yichen Fang, Jiadi Zhu, Keqin Liu, Liying
  Xu, Yimao Cai, Xiaoqin Yan, Yuchao Yang, and Ru~Huang.
\newblock {Bipolar to unipolar mode transition and imitation of metaplasticity
  in oxide based memristors with enhanced ionic conductivity}.
\newblock \emph{Journal of Applied Physics}, 124\penalty0 (15):\penalty0
  152103, oct 2018.
\newblock ISSN 10897550.
\newblock \doi{10.1063/1.5037962}.
\newblock URL \url{http://aip.scitation.org/doi/10.1063/1.5037962}.

\bibitem[Liu et~al.(2018)Liu, Liu, Chiu, Di, Wu, Wang, Hou, and Lai]{Liu2018}
Bo~Liu, Zhiwei Liu, In~Shiang Chiu, MengFu Di, YongRen Wu, Jer~Chyi Wang,
  Tuo~Hung Hou, and Chao~Sung Lai.
\newblock {Programmable Synaptic Metaplasticity and below Femtojoule Spiking
  Energy Realized in Graphene-Based Neuromorphic Memristor}.
\newblock \emph{ACS Applied Materials and Interfaces}, 10\penalty0
  (24):\penalty0 20237--20243, 2018.
\newblock ISSN 19448252.
\newblock \doi{10.1021/acsami.8b04685}.
\newblock URL \url{https://pubs.acs.org/doi/10.1021/acsami.8b04685}.

\bibitem[Lee et~al.(2018)Lee, Hwang, Woo, Kim, Kim, and Nahm]{Lee2018}
Tae~Ho Lee, Hyun~Gyu Hwang, Jong~Un Woo, Dae~Hyeon Kim, Tae~Wook Kim, and Sahn
  Nahm.
\newblock {Synaptic Plasticity and Metaplasticity of Biological Synapse
  Realized in a KNbO3 Memristor for Application to Artificial Synapse}.
\newblock \emph{ACS Applied Materials and Interfaces}, 10\penalty0
  (30):\penalty0 25673--25682, 2018.
\newblock ISSN 19448252.
\newblock \doi{10.1021/acsami.8b04550}.
\newblock URL \url{www.acsami.org}.

\bibitem[Chang et~al.(2011)Chang, Jo, and Lu]{Chang2011}
Ting Chang, Sung~Hyun Jo, and Wei Lu.
\newblock {Short-term memory to long-term memory transition in a nanoscale
  memristor}.
\newblock \emph{ACS Nano}, 5\penalty0 (9):\penalty0 7669--7676, 2011.
\newblock ISSN 19360851.
\newblock \doi{10.1021/nn202983n}.
\newblock URL \url{www.acsnano.org}.

\bibitem[Tan et~al.(2016)Tan, Yang, Terabe, Yin, Zhang, and Guo]{Tan2016}
Zheng~Hua Tan, Rui Yang, Kazuya Terabe, Xue~Bi Yin, Xiao~Dong Zhang, and Xin
  Guo.
\newblock {Synaptic Metaplasticity Realized in Oxide Memristive Devices}.
\newblock \emph{Advanced Materials}, 28\penalty0 (2):\penalty0 377--384, jan
  2016.
\newblock ISSN 15214095.
\newblock \doi{10.1002/adma.201503575}.
\newblock URL \url{http://doi.wiley.com/10.1002/adma.201503575}.

\bibitem[Ohno et~al.(2011)Ohno, Hasegawa, Tsuruoka, Terabe, Gimzewski, and
  Aono]{Ohno2011}
Takeo Ohno, Tsuyoshi Hasegawa, Tohru Tsuruoka, Kazuya Terabe, James~K
  Gimzewski, and Masakazu Aono.
\newblock {Short-term plasticity and long-term potentiation mimicked in single
  inorganic synapses}.
\newblock \emph{Nature Materials}, 10\penalty0 (8):\penalty0 591--595, 2011.
\newblock ISSN 14764660.
\newblock \doi{10.1038/nmat3054}.
\newblock URL \url{www.nature.com/naturematerials}.

\bibitem[Laborieux et~al.(2021)Laborieux, Ernoult, Hirtzlin, and
  Querlioz]{Laborieux2021}
Axel Laborieux, Maxence Ernoult, Tifenn Hirtzlin, and Damien Querlioz.
\newblock {Synaptic metaplasticity in binarized neural networks}.
\newblock \emph{Nature Communications}, 12\penalty0 (1):\penalty0 2549, dec
  2021.
\newblock ISSN 2041-1723.
\newblock \doi{10.1038/s41467-021-22768-y}.
\newblock URL \url{https://doi.org/10.1038/s41467-021-22768-y
  http://www.nature.com/articles/s41467-021-22768-y}.

\bibitem[Giotis et~al.(2020{\natexlab{a}})Giotis, Serb, Stathopoulos, Michalas,
  Khiat, and Prodromakis]{Giotis2020}
Christos Giotis, Alex Serb, Spyros Stathopoulos, Loukas Michalas, Ali Khiat,
  and Themis Prodromakis.
\newblock {Bidirectional Volatile Signatures of Metal-Oxide Memristors-Part I:
  Characterization}.
\newblock \emph{IEEE Transactions on Electron Devices}, 67\penalty0
  (11):\penalty0 5158--5165, nov 2020{\natexlab{a}}.
\newblock ISSN 15579646.
\newblock \doi{10.1109/TED.2020.3014854}.

\bibitem[Giotis et~al.(2020{\natexlab{b}})Giotis, Serb, Stathopoulos, and
  Prodromakis]{Giotis2020b}
C.~Giotis, A.~Serb, S.~Stathopoulos, and T.~Prodromakis.
\newblock {Bidirectional Volatile Signatures of Metal-Oxide Memristors-Part II:
  Modeling}.
\newblock \emph{IEEE Transactions on Electron Devices}, 67\penalty0
  (11):\penalty0 5166--5173, nov 2020{\natexlab{b}}.
\newblock ISSN 15579646.
\newblock \doi{10.1109/TED.2020.3022343}.

\bibitem[Bhalla(2014)]{Bhalla2014}
Upinder~S. Bhalla.
\newblock {Molecular computation in neurons: A modeling perspective}.
\newblock \emph{Current Opinion in Neurobiology}, 25:\penalty0 31--37, apr
  2014.
\newblock ISSN 09594388.
\newblock \doi{10.1016/j.conb.2013.11.006}.

\bibitem[Messaris et~al.(2018)Messaris, Serb, Stathopoulos, Khiat, Nikolaidis,
  and Prodromakis]{Messaris2018}
Ioannis Messaris, Alexander Serb, Spyros Stathopoulos, Ali Khiat, Spyridon
  Nikolaidis, and Themistoklis Prodromakis.
\newblock {A Data-Driven Verilog-A ReRAM Model}.
\newblock \emph{IEEE Transactions on Computer-Aided Design of Integrated
  Circuits and Systems}, 37\penalty0 (12):\penalty0 3151--3162, 2018.
\newblock \doi{10.1109/tcad.2018.2791468}.

\bibitem[Fusi and Abbott(2007)]{Fusi2007}
S~Fusi and L~F Abbott.
\newblock {Limits on the memory storage capacity of bounded synapses}.
\newblock \emph{Nat Neurosci}, 10\penalty0 (4):\penalty0 485--493, 2007.
\newblock \doi{10.1038/nn1859}.
\newblock URL \url{https://www.ncbi.nlm.nih.gov/pubmed/17351638}.

\bibitem[Benna and Fusi(2017)]{Benna2017}
M~K Benna and S~Fusi.
\newblock {Efficient online learning with low-precision synaptic variables}.
\newblock \emph{2017 Fifty-First Asilomar Conference on Signals, Systems, and
  Computers}, pages 1610--1614, 2017.

\bibitem[Matthey et~al.(2015)Matthey, Bays, and Dayan]{Matthey2015}
Loic Matthey, Paul~M. Bays, and Peter Dayan.
\newblock {A Probabilistic Palimpsest Model of Visual Short-term Memory}.
\newblock \emph{PLoS Computational Biology}, 11\penalty0 (1):\penalty0 1004003,
  jan 2015.
\newblock ISSN 15537358.
\newblock \doi{10.1371/journal.pcbi.1004003}.
\newblock URL \url{http://www.cns.nyu.edu/malab/}.

\bibitem[Luck and Vogel(1997)]{Luck1997}
Steven~J. Luck and Edward~K. Vogel.
\newblock {The capacity of visual working memory for features and
  conjunctions}.
\newblock \emph{Nature}, 390\penalty0 (6657):\penalty0 279--284, nov 1997.
\newblock ISSN 00280836.
\newblock \doi{10.1038/36846}.
\newblock URL \url{https://www.nature.com/articles/36846}.

\bibitem[Zhang and Luck(2008)]{Zhang2008}
Weiwei Zhang and Steven~J. Luck.
\newblock {Discrete fixed-resolution representations in visual working memory}.
\newblock \emph{Nature}, 453\penalty0 (7192):\penalty0 233--235, may 2008.
\newblock ISSN 14764687.
\newblock \doi{10.1038/nature06860}.
\newblock URL \url{www.nature.com/nature.}

\bibitem[Lisman et~al.(2002)Lisman, Schulman, and Cline]{Lisman2002}
John Lisman, Howard Schulman, and Hollis Cline.
\newblock {The molecular basis of CaMKII function in synaptic and behavioural
  memory}.
\newblock \emph{Nature Reviews Neuroscience}, 3\penalty0 (3):\penalty0
  175--190, 2002.
\newblock ISSN 14710048.
\newblock \doi{10.1038/nrn753}.
\newblock URL \url{https://www.nature.com/articles/nrn753}.

\bibitem[Miller et~al.(2005)Miller, Zhabotinsky, Lisman, and Wang]{Miller2005}
Paul Miller, Anatol~M Zhabotinsky, John~E Lisman, and Xiao~Jing Wang.
\newblock {The stability of a stochastic CaMKII switch: Dependence on the
  number of enzyme molecules and protein turnover}.
\newblock \emph{PLoS Biology}, 3\penalty0 (4):\penalty0 0705--0717, 2005.
\newblock ISSN 15449173.
\newblock \doi{10.1371/journal.pbio.0030107}.
\newblock URL \url{www.plosbiology.org}.

\bibitem[{Le Gallo} et~al.(2018){Le Gallo}, Sebastian, Mathis, Manica, Giefers,
  Tuma, Bekas, Curioni, and Eleftheriou]{LeGallo2018}
Manuel {Le Gallo}, Abu Sebastian, Roland Mathis, Matteo Manica, Heiner Giefers,
  Tomas Tuma, Costas Bekas, Alessandro Curioni, and Evangelos Eleftheriou.
\newblock {Mixed-precision in-memory computing}.
\newblock \emph{Nature Electronics}, 1\penalty0 (4):\penalty0 246--253, apr
  2018.
\newblock ISSN 25201131.
\newblock \doi{10.1038/s41928-018-0054-8}.
\newblock URL \url{https://doi.org/10.1038/s41928-018-0054-8}.

\bibitem[Hu et~al.(2015)Hu, Liu, Liu, Chen, Wang, Yu, Deng, Yin, and
  Hosaka]{Hu2015}
S.~G. Hu, Y.~Liu, Z.~Liu, T.~P. Chen, J.~J. Wang, Q.~Yu, L.~J. Deng, Y.~Yin,
  and Sumio Hosaka.
\newblock {Associative memory realized by a reconfigurable memristive Hopfield
  neural network}.
\newblock \emph{Nature Communications}, 6, 2015.
\newblock ISSN 20411723.
\newblock \doi{10.1038/ncomms8522}.
\newblock URL \url{www.nature.com/naturecommunications}.

\bibitem[Vaswani et~al.(2017)Vaswani, Shazeer, Parmar, Uszkoreit, Jones, Gomez,
  Kaiser, and Polosukhin]{Vaswani2017}
Ashish Vaswani, Noam Shazeer, Niki Parmar, Jakob Uszkoreit, Llion Jones,
  Aidan~N Gomez, {\L}ukasz Kaiser, and Illia Polosukhin.
\newblock {Attention is all you need}.
\newblock In \emph{Advances in Neural Information Processing Systems}, volume
  2017-Decem, pages 5999--6009, 2017.

\bibitem[Savin et~al.(2014)Savin, Dayan, and Lengyel]{Savin2014}
C~Savin, P~Dayan, and M~Lengyel.
\newblock {Optimal recall from bounded metaplastic synapses: predicting
  functional adaptations in hippocampal area CA3}.
\newblock \emph{PLoS Comput Biol}, 10\penalty0 (2):\penalty0 e1003489, 2014.
\newblock \doi{10.1371/journal.pcbi.1003489}.
\newblock URL \url{https://www.ncbi.nlm.nih.gov/pubmed/24586137}.

\bibitem[Covi et~al.(2019)Covi, Ielmini, Lin, Wang, Stecconi, Milo, Bricalli,
  Ambrosi, Pedretti, and Tseng]{Covi2019}
E.~Covi, D.~Ielmini, Y.~H. Lin, W.~Wang, T.~Stecconi, V.~Milo, A.~Bricalli,
  E.~Ambrosi, G.~Pedretti, and T.~Y. Tseng.
\newblock {A volatile RRAM synapse for neuromorphic computing}.
\newblock In \emph{2019 26th IEEE International Conference on Electronics,
  Circuits and Systems, ICECS 2019}, pages 903--906. Institute of Electrical
  and Electronics Engineers Inc., nov 2019.
\newblock ISBN 9781728109961.
\newblock \doi{10.1109/ICECS46596.2019.8965044}.

\end{thebibliography}

%%% Uncomment this section and comment out the \bibliography{references} line above to use inline references.
% \begin{thebibliography}{1}

% 	\bibitem{kour2014real}
% 	George Kour and Raid Saabne.
% 	\newblock Real-time segmentation of on-line handwritten arabic script.
% 	\newblock In {\em Frontiers in Handwriting Recognition (ICFHR), 2014 14th
% 			International Conference on}, pages 417--422. IEEE, 2014.

% 	\bibitem{kour2014fast}
% 	George Kour and Raid Saabne.
% 	\newblock Fast classification of handwritten on-line arabic characters.
% 	\newblock In {\em Soft Computing and Pattern Recognition (SoCPaR), 2014 6th
% 			International Conference of}, pages 312--318. IEEE, 2014.

% 	\bibitem{hadash2018estimate}
% 	Guy Hadash, Einat Kermany, Boaz Carmeli, Ofer Lavi, George Kour, and Alon
% 	Jacovi.
% 	\newblock Estimate and replace: A novel approach to integrating deep neural
% 	networks with existing applications.
% 	\newblock {\em arXiv preprint arXiv:1804.09028}, 2018.

% \end{thebibliography}

\end{document}